\documentclass[twocolumn,showpacs,preprintnumbers,amsmath,amssymb,prl,superscriptaddress]{revtex4-1}
\usepackage[T1]{fontenc}
\usepackage[latin1]{inputenc}
\usepackage{graphicx}

\expandafter\let\csname equation*\endcsname\relax
\expandafter\let\csname endequation*\endcsname\relax
\usepackage{amsmath}

\newcommand{\etal}{\mbox{\emph{et al.}\hspace{-1pt}}~}

\newcommand{\ev}{\mathrm{eV}}

\newcommand{\angstrom}{\textup{\AA}}
\newcommand{\nm}{\mathrm{nm}}
\newcommand{\cm}{\mathrm{cm}}

\newcommand{\Eqref}[1]{Eq.~(\ref{#1})}
\newcommand{\Figref}[1]{Fig.~\ref{#1}}

\makeatother

\newcommand{\siesta}{\textsc{Siesta}}
\newcommand{\tsiesta}{\textsc{TranSiesta}}
\newcommand{\tbtrans}{\textsc{TBtrans}}
\newcommand{\sisl}{\textsc{sisl}}
\newcommand{\kwant}{\textsc{kwant}}

\makeatother

\newif\ifchanged
\changedtrue
\changedfalse

\newcommand{\change}[1]{\ifchanged\textcolor{red}{#1}\else#1\fi}

\begin{document}

\title[Large-scale tight-binding simulations of quantum transport in ballistic graphene]{Large-scale tight-binding simulations of quantum transport in ballistic graphene}

\author{Gaetano Calogero}
\author{Nick R. Papior}
\author{Peter B{\o}ggild}
\author{Mads Brandbyge}

\affiliation{Dept. of Micro- and Nanotechnology, Technical University of Denmark, Center for Nanostructured Graphene (CNG), 
	{\O}rsteds Plads, Bldg.~345E, DK-2800 Kongens Lyngby, Denmark}
\email{gaca@nanotech.dtu.dk}

\date{\today}

\begin{abstract}
Graphene has proven to host outstanding mesoscopic effects involving massless Dirac
quasiparticles travelling ballistically resulting in the current flow exhibiting light-like
behaviour. A new branch of 2D electronics inspired by the standard principles of optics
is rapidly evolving, calling for a deeper understanding of transport in large-scale
devices at a quantum level. Here we perform large-scale quantum transport calculations
based on a tight-binding model of graphene and the non-equilibrium Green's function
method and include the effects of $p-n$ junctions of different shape, magnetic field, and
absorptive regions acting as drains for current. We stress the importance of choosing
absorbing boundary conditions in the calculations to correctly capture how current flows
in the limit of infinite devices. As a specific application we present a fully quantum-mechanical
framework for the ``2D Dirac fermion microscope'' recently proposed by B{\o}ggild \etal\ [Nat. Comm. 8, 10.1038 (2017)], tackling several key electron-optical effects therein
predicted via semiclassical trajectory simulations, such as electron beam collimation,
deflection and scattering off Veselago dots. Our results confirm that a semiclassical
approach to a large extend is sufficient to capture the main transport features in the
mesoscopic limit and the optical regime, but also that a richer electron-optical
landscape is to be expected when coherence or other purely quantum effects
are accounted for in the simulations.
\end{abstract}

\maketitle

\section{Introduction}

Graphene has proven to be the scene of unprecedented mesoscopic effects, hosting massless
Dirac quasiparticles that travel with little scattering. These relativistic charge
carriers can move ballistically across $\mu$m-long distances at room temperature
\cite{Taychatanapat2013}, so far reaching mean free paths of the order 30 $\mu$m at low
temperatures \cite{Banszerus2016}. They can undergo negative refraction when passing
$p-n$ junctions \cite{Cheianov2007} and can be manipulated by external electromagnetic
fields \cite{Chen2016}, thus being easily emitted, collimated, steered or focused like
rays of light. As a result a new type of 2D electronics complying with the principles of
optics is rapidly finding its way in the 2D materials community, supported by the relentless
progress towards large-scale production of high-quality graphene.

\begin{figure}
	\centering
	\includegraphics[width=0.95\columnwidth]{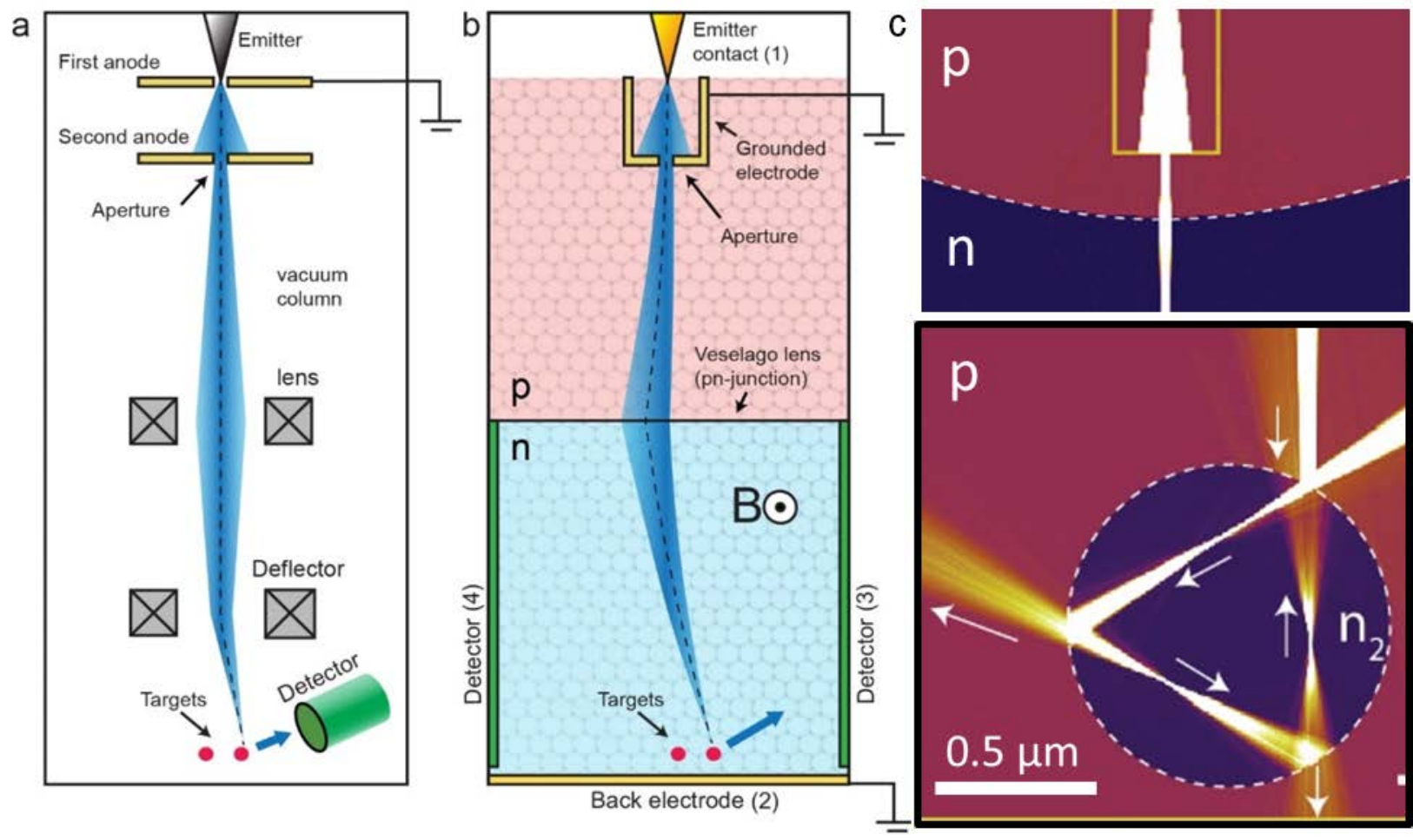}
	\caption{(a) Schematics of a typical electron microscope, where electrons are injected, collimated, deflected and detected inside a 3D vacuum chamber. (b) Schematics of a 2D Dirac fermion microscope, where electrons move in straight trajectories within 2D graphene in analogy to electrons traveling inside a 3D vacuum chamber. By using $p-n$ junctions as tunable lenses and magnetic fields for deflection, electrons can be injected, collimated, directed and focused onto various targets. (c) Snapshots of simulated semiclassical trajectories of electrons, showing injection into graphene, collimation via a grounded aperture and a parabolic $p-n$ junction, scattering of the collimated beam off a Veselago dot. Figures are adapted from B{\o}ggild \etal \cite{Boggild2017}, to which the reader can refer for further details.}
	\label{MonteCarlo}
\end{figure}

One example, recently brough up by B{\o}ggild and co-authors \cite{Boggild2017}, is to
combine different graphene-based electron-optics components in a ``2D Dirac fermion
microscope'' (DFM) as in \Figref{MonteCarlo}. Here electron emitters/guns, collimating
apertures \cite{Barnard2017}, tunable lenses, deflectors, and detectors are imagined to be
incorporated in a graphene ``vacuum chamber'' to image different types of targets, such as
metal-graphene interfaces, grain boundaries, edges, defects, adsorbed molecules,
nanoparticles, quantum dots, or plasmonic superstructures. The authors provide a
perspective view on how such a tool can be realistically implemented and operated,
proposing practical architectures and design rules for all of its 2D components, based on
state-of-the-art achievements in graphene technology.  
%While the image resolution of a DFM can never rival those of established electron or scanning probe microscopies, the DFM embodies a versatile electron transport measurement system which circumvents the edge scattering, inflexibility and limitations of a permanent, massive experimental apparatus.\MB{Note sure what exactly you mean.. but maybe Peter can elaborate here?}

The approach used by B{\o}ggild \etal\ for large-scale electron transport simulations is
purely semiclassical and belongs to a broadly used class of simulation known as
\emph{billiard models}. In the last few years these models have proven to successfully
provide insights on the overall magneto-transport characteristics of graphene
\cite{Miao2007, Lackner2013, Taychatanapat2015, Wurm2011} and other large-scale ballistic
devices in the mesoscopic limit \cite{Molenkamp1990}.

However, despite allowing computation with little time and memory consumption,
semiclassical transport simulations always need to be calibrated with measured macroscopic
parameters, such as mobility or diffusion coefficients. Most importantly, quantum effects
such as coherence are not naturally included in semiclassical simulations, despite their
importance for describing phenomena such as magnetic focusing, chiral tunneling in the
ballistic regime or conductance fluctuations in the diffusive regime
\cite{Caridad2016,Jiang2017}. Diffraction is also expected to have major implications in
devices where Dirac fermions pass through apertures smaller than their Fermi wavelength
$\lambda_{\mathrm F}$ or scatter off small objects. Future realization and operation of
complex relativistic electron-optics graphene systems thus calls for a deeper
understanding of transport at the quantum level, where the full quantum nature of Dirac
fermions is accounted for. 

Along this line it is also decisive to be able to access
simulations at the scale of experimental devices, which often range from hundreds of
nano-meters to a few microns. At the same time it is crucial to provide reliable
benchmarks to measurements in the presence of defects, interfaces, or disorder, where
details matter on the atomic scale. The huge number of atoms contained in the typical
experimental systems prohibits the application of usual \emph{ab-initio} electronic
structure techniques like density functional theory (DFT), where the detailed
quantum-chemical structure of every atom is taken into account. Hence, the development of
novel high-performance computational methods to enable quantum transport simulations at
experimentally relevant device dimensions is essential.

Here we perform quantum transport calculations of large tight-binding (TB) models of
graphene using the non-equilibrium Green's function method (NEGF). We will report on the
multi-functionality and performance of our tools while studying transport in large
graphene flakes on the scale of hundreds of $\nm$ in the presence of $p-n$ junctions, magnetic
field and/or absorptive regions. Our main focus will be to reproduce from a fully atomistic
perspective some key features of electron transport in a DFM, such as electron beam
collimation, deflection, and scattering off circular Veselago dots (VD)
\cite{Cheianov2007}. We will emphasize how different choices of boundary conditions lead
to different density patterns, providing a simple computational solution to minimize the
occurrence of artificial features in the current, e.g. using hard-wall or periodic
boundaries.  The manuscript is organized as follows. We will first briefly review the
state of the art of the computational methods that can be used for transport simulations
of large-scale graphene devices. This is followed by an overview of the used methods, the
code, and the setup we use to carry out our calculations. We will then discuss the effects
of adopting different boundary conditions in the device. To conclude we will present a direct comparison between our results and those
reported in \cite{Boggild2017}, highlighting similarities and differences between the
semiclassical and quantum simulations.

\subsection{Methods for large-scale graphene transport simulations}

For any quantum transport technique to efficiently address realistic graphene devices, it
is vital to describe the underlying electronic structure in a computationally efficient
manner.  A broadly adopted solution is to model the electronic structure of graphene using
the tight-binding approximation \cite{CastroNeto2009, FoaTorres, Meunier2016}. This
approximation is in its ``cheapest'' textbook version, where the Hamiltonian is
orthogonal and includes only interactions between nearest-neighbor $\pi$ ($p_z$) orbitals,
able to capture the main qualitative features of the graphene band structure. Due to its
versatility, this method has often been at the center of new improvements and
developments. In particular we point out the simple scaling approach proposed by Liu \etal
\cite{Liu2012,Liu2015}, who provide a simple condition to obtain band structure
invariance, while simultaneously adjusting the lattice constant and the hopping parameter
in tight-binding models of graphene. A similar approach is the one
suggested by Beconcini \etal\ \cite{Beconcini2016}, who manage to achieve band structure
invariance by using the Fermi energy as key scaling parameter, 
%thus generalizing Liu's approach to multi-terminal devices and nonlocal transport measurements
thus accessing simulations of multi-terminal devices and nonlocal transport measurements. In more detail, they show that a geometrical downward
scaling of the system size can be accompanied by an upward scaling of the Fermi energy
in such a way that the number of electronic states responsible for transport is kept
constant. Both of these scaling approaches have proven to be very efficient tools to
interpret magnetic focusing experiments involving micron-size multi-terminal devices
\cite{Petrovic2017, Lagasse2017}, where factors such as edges, chemical functionalization,
structural disorder or contact with metal do not disrupt the relevant transport features.

Electron transport for very large system dimensions can then be achieved by coupling a TB
Hamiltonian with different quantum transport formalisms \cite{Uppstu2014}. For example, 
%the Ehrenfest Quantum Molecular Dynamics method \cite{Tong2013} can give a full time-dependent quantum mechanical treatment of non-equilibrium electrons states while regarding the ions as classical particles. 
hybrid Monte Carlo algorithms on lattice \cite{Buividovich2012} or
Wave-Packet Dynamics \cite{Mark2017, Vancso2014, Krueckl2009, Haefner2011} have been
used to study large-scale transport in graphene. The latter in particular is a very
intuitive method which has the advantage of giving direct access to the real-space and real-time electron wave-packet propagation over a graphene lattice. Here energy resolution can be
obtained by Fourier transforming in time domain, at the cost of including a very fine time
discretization.  The majority of the other transport formalisms are based on the
Landauer-B\"uttiker theory. The Kubo-Greenwood formalism is a very popular example
\cite{Ortmann2011,Lherbier2012,BarriosVargas2017,Calderin2017,Ervasti2015}, which turns
out to be an ideal choice when studying diffusive large-scale graphene systems described
by a mobility or conductivity. Along similar lines the patched Green's function technique
can be used to introduce open-boundary self-energy terms in the device Hamiltonian to
describe its connection to an infinite sample \cite{Settnes2015}. 
Complementary to this the TB-NEGF method
is a popular choice for ballistic transport \cite{Datta2000} where the target is most
often conductance including explicit descriptions of multiple electrodes, rather than
conductivity. It allows for self-consistent mean-field description of the potential,
e.g. using a Hubbard-type model\cite{Hancock2010} including the possibility to study the
effect of spin-polarization. Many simulation packages are now implementing the NEGF
scheme as a standard feature to calculate electron transport in nanostructures, enabling
its application to tight-binding Hamiltonians as well as other electronic structure models
with higher level of accuracy such as DFT \cite{Brandbyge2002} which can scale linearly
with length of the system in the transport direction using recursive Green's function
methods.

\section{Computational tools}

\change{Our quantum transport simulations are based on the NEGF method and a nearest-neighbour TB Hamiltonian using the standard expressions of transmission in terms of the retarded Green's function \cite{Datta1997, Datta2000}}.
\change{For the particular implementation we use the open-source} \tbtrans\ and \sisl\ \cite{Papior2017,Papior2017a} tools distributed with the
\tsiesta\ software package \cite{Papior2017}.  \tsiesta\ is a tool for high-performance
DFT+NEGF self-consistent calculations. It relies on advanced matrix inversion
algorithms to efficiently obtain the Green's functions for large, multi-terminal systems
at various electrostatic conditions (e.g. gating \cite{Papior2015}), while the
charge-density is obtained using contour integration of the spectral densities.

\tbtrans\ is a ``post-processing'' NEGF code which provides a flexible interface to DFT as
well as user-defined tight-binding Hamiltonians, using Python as back-end. It enables
large-scale tight-binding transport calculations of spectral physical quantities,
interpolated $I$-$V$ curves, transmission eigenchannels \cite{Papior2017} and/or
orbital/bond-currents for setups that can easily exceed millions of orbitals on few-core
machines. The possibility of using Bloch expansion for electrodes with periodicity
transverse to the transport direction and customizing the effective shape of the device
region makes it possible to increase the scale of transmission calculations even
further\cite{PapiorThesis}.  Complementary to \tbtrans\ \sisl\ was developed as a Python
package to create and manipulate large-scale (non-)orthogonal tight-binding models for
arbitrary geometries, with any number of orbitals and any periodicity. It allows to read
external Hamiltonians and real-space grids from various DFT programs (e.g. \siesta\
\cite{Soler2002} or Wannier90 \cite{Mostofi2014}), providing user-friendly routines for
post-processing and output of electronic structure and transport calculations.  To include
the effects of doping, magnetic field or absorptive potentials in our system we exploit
the \tbtrans\ capability of customizing the device Green's function via real
(complex) $\mathbf k$- (energy-) dependent $\delta\mathbf H(E, \mathbf k)$ perturbative
terms:
\begin{equation}
\begin{split}
\mathbf G(E, \mathbf k) &= \left[ \mathbf S_0(\mathbf k) (E + i\eta) - \mathbf H_0(E, \mathbf k) \vphantom{\sum_i\boldsymbol\Sigma_i}\right. \\ & \left. - \sum_i\boldsymbol\Sigma_i(E,\mathbf k) - \delta\mathbf H(E, \mathbf k) \right]^{-1}
\end{split}
\end{equation}
%\begin{equation}
%\mathbf G(E, \mathbf k) = \left[ \mathbf S_0(\mathbf k) (E + i\eta) - \mathbf H_0(E, \mathbf k) - \sum_i\boldsymbol\Sigma_i(E,\mathbf k) - \delta\mathbf H(E, \mathbf k) \right]^{-1}
%\end{equation}

Here $\mathbf S_0$ and $\mathbf H_0$ are the unperturbed overlap and Hamiltonian in the
device region, while $\boldsymbol\Sigma_i$ is the self-energy for each semi-infinite
electrode $i$. 
\change{If we call $\delta\mathbf H_{p-n}$, $\delta\mathbf H_{CAP}$ and $\delta\mathbf H_{B}$ the $\mathbf k$- and energy independent perturbations to the Hamiltonian caused by $p-n$ junctions, absorptive regions (like those simulated using complex absorbing potential (CAP) \cite{Xie2014,Yu2015}) and magnetic field $B$, respectively, we can incorporate the effects of these mechanisms in the expression for $\delta\mathbf H$ as:}
\begin{equation}
	\delta\mathbf H = \delta\mathbf H_{p-n} + \delta\mathbf H_{CAP} + \delta\mathbf H_{B}
	\label{dH}
\end{equation}
In the following sections we provide details on \change{the physics behind the terms} in \Eqref{dH} and how we construct them \change{in the TB calculation}. 

We use \sisl\ to set up a nearest-neighbor orthogonal TB Hamiltonian for a two-probe
graphene device, with carbon-carbon bond length $a_0=0.142\,\nm$ and hopping parameter
$t_0=2.7\, \ev$. Our supercell, inspired by the hetero-dimensional graphene junctions
studied in \cite{Wang2010}, consists of a $2\,\nm$ wide zigzag graphene nanoribbon acting as
point-like source electrode at the edge of a $100\,\nm\times 100\,\nm$ graphene flake with
$395.940$ atoms (sites). 
%\footnote{Armchair ribbons have been also investigated. Although momentum injection in the flake has a different anisotropy, the use of a CAP aperture does not change the results obtained for DFM calculations.}. We inject current through the nanoribbon and collect it in a wide drain lead located at the opposite edge.

\subsection{p-n junctions}

\begin{figure}
	\centering
	\includegraphics[width=1.0\columnwidth]{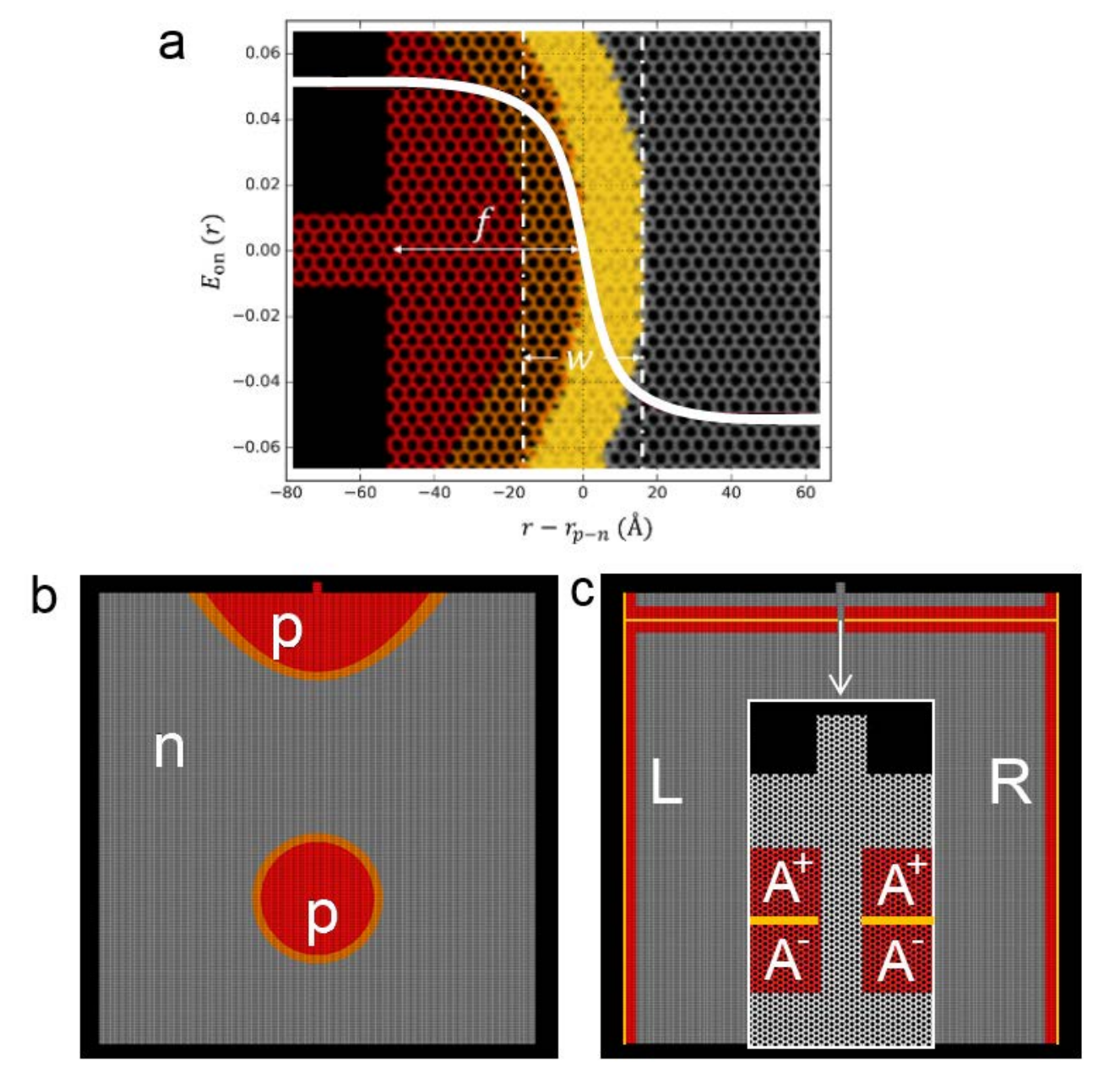}
	\caption{
		(a) Geometry of the graphene device
		considered here where $p$-doped (red) parabolic and circular regions are created on a
		$n$-doped (grey) graphene flake.
		(b) General Fermi-like profile of the smooth parabolic $p-n$ junction. On-site energy along an axis passing through the parabola's
		vertex is shown as a function of distance between the ribbon/flake interface and the
		junction average position. Red (grey) sites in the overlaid ribbon/flake geometry
		have $E_{\mathrm{on}}=+0.05\,\ev$ ($E_{\mathrm{on}}=-0.05\,\ev$), whereas all other sites have
		gradually varying $E_{\mathrm{on}}$ according to \Eqref{eqpn}.
		(c) Sites in geometry equipped with complex absorbing potential (CAP) are shown
		in red. Absorption occurs gradually starting from $\Delta z\approx30\angstrom$ from
		the yellow lines, where maximum absorption takes place due to the singularity in
		\Eqref{fz}. 
	}
	
	% \caption{\MB{Put a,b,c labels on the figures} (a) Geometry of the graphene device
	%   considered here. All sites equipped with complex absorbing potential (CAP) are shown
	%   in red. Absorption occurs gradually starting from $\Delta z\approx30\angstrom$ from
	%   the yellow lines, where maximum absorption takes place due to the singularity in
	%   \Eqref{fz}. (b) General Fermi-like profile of the smooth parabolic p-n junction used
	%   in our calculations. On-site energy along a axis passing through the parabola's
	%   vertex is shown as a function of distance between the ribbon/flake interface and the
	%   junction average position. Red (grey) sites in the overlaid ribbon/flake geometry
	%   have $E_{on}=+0.05\,\ev$ ($E_{on}=-0.05\,\ev$), whereas all other sites have
	%   gradually varying $E_{on}$ according to \Eqref{eqpn}. (c) Example of full device
	%   configuration where $p$-doped (red) parabolic and circular regions are created on a
	%   $n$-doped (grey) graphene flake.}
	
	\label{LRB}
\end{figure}

In an orthogonal tight-binding model doping is straightforward to introduce via a
global or local adjustment of the diagonal (on-site) elements of the Hamiltonian. \change{In our calculations we generate smooth symmetric $p-n$ junctions using:}
\begin{equation}
	\delta\mathbf H_{p-n}=
	\begin{bmatrix}
		\Delta E/2 \cdot \delta_{ij} & 0 & 0 \\
		0 & E_{\mathrm{on}}(r) \cdot \delta_{ij} & 0 \\
		0 & 0 & - \Delta E/2 \cdot \delta_{ij}
	\end{bmatrix}
\end{equation}
\change{where the junction potential profile is given by}
\begin{equation}
	E_{\mathrm{on}}(r)=\frac{\Delta E}{2}\left[ \frac{2}{1+\exp[-\alpha(r-r_{p-n})/w]}-1\right]
	\label{eqpn}
\end{equation}
\change{and the junction thickness $w$ is set to $w\approx2\,\nm$.}
The junction profile $r_{p-n}$ can be chosen to be linear or can conveniently be
shaped to achieve Veselago lensing of electrons \cite{Chen2016}. In particular, as
suggested by Liu et al. \change{\cite{Liu2017}}, electrons can be focused into a sharp collimated
beam using a parabolic $p-n$ junction with focal point located at the point source. In our
model for the DFM source we thus make use of such shape, placing the parabola at a focal distance $f\approx 4\,\nm$ from the injection point, 
as illustrated in \Figref{LRB}a and b. In the same figure we show that
this method also can be used to construct circular p-n juctions, or Veselago dots (VD)
\cite{Cserti2007}, here used as targets for the collimated beam of electrons.

\subsection{Complex absorbing potential (CAP)}

In standard NEGF transport problems one usually deals with infinite open systems, where
semi-infinite leads are included in the calculation of the Green's function via a
self-energy in the device Hamiltonian. For large geometries a major computational
limitation is the number and the size of semi-infinite leads. This limitation can be
efficiently overcome by replacing the Hamiltonian of the leads with complex absorbing
potentials (CAP) that completely absorb the incident wave-function \cite{Xie2014,
	Yu2015}. This method allows reduction of the original open system to a finite closed
system without disrupting current conservation. For example, the Hamiltonian matrix for a
two-probe system where CAP is used to replace the electrode self-energies can be written
as the sum of the standard two-probe Hamiltonian matrix plus a purely imaginary
$\delta\mathbf H_{CAP}$ term added to the diagonal matrix elements:
\begin{equation}
	\delta\mathbf H_{CAP}=
	\begin{bmatrix}
		-i W_L(r) \cdot \delta_{ij} & 0 & 0 \\
		0 & 0 & 0 \\
		0 & 0 & -i W_R(r) \cdot \delta_{ij}
	\end{bmatrix}.
\end{equation}
Here we assume transport to occur along the $r$ direction, and define $W_L(r)$ and $W_R(r)$ as
\begin{equation}
	W_{L/R}(r)=\frac{\hbar^2}{2m}\left( \frac{2\pi}{\Delta r} \right)^2 f(r)
	\label{CAP}
\end{equation}
where $f(r)$ is a smooth function of the form 
\begin{equation}
	f(r)=\frac{4}{c^2}\left[\left(\frac{\Delta r}{r_f-2r_i+r}\right)^2+\left(\frac{\Delta r}{r_f-r}\right)^2-2\right]
	\label{fz}
\end{equation}
Here $r_i$ and $r_f$ are the starting and ending points of the CAP region in the device,
respectively, $\Delta r=r_f-r_i$ is its length and $c$ is a constant numerical parameter set to be equal to
2.62 \cite{Yu2015}. Note how $f(r)$ diverges as $r$ tends
to $r_f$, turning the semi-infinite lead into a finite lead. The CAP expression in
\Eqref{CAP} is based on purely semiclassical arguments, and may lead to reflection at
$r_f$, which reduces the final transmission. However this can be mitigated by increasing
the length $\Delta r$ of the CAP region, until the spectra are in agreement with those
obtained in the original open system. In order to estimate a suitable value for the
thickness of CAP regions in our device configuration, we have calculated transmission
using semi-infinite source and drain electrodes along $\pm y$ and applying periodic
boundary conditions (PBC) along the transverse direction $x$. We then compared this with
the transmission obtained by setting CAP rather than PBC on the cell boundaries along
$x$. We find that an almost exact overlap between the two spectra is achieved by setting
$\Delta r\ge0.3\,\nm$. 

CAP can also be used to design narrow absorbing areas such as those generated by contacts
with grounded electrodes \cite{Barnard2017}. A proper comparison with simulations of a DFM reported in
Ref.~\cite{Boggild2017} requires an isotropic point-like source of electrons. However, a
hetero-dimensional graphene junction, such as the ribbon considered here, is known to
produce isotropic injection only for electron energies very close to the Dirac point,
whereas preferential injection takes over at higher energies, with angles depending on the
ribbon symmetry \cite{Wang2010}. In order to ensure isotropic injection we therefore place
an absorptive pinhole with an opening of $1.5\,\nm$ at a distance of $3\,\nm$ from the
ribbon/flake interface, mimicking apertures generated by grounded electrodes
\cite{Barnard2017}. In \Figref{LRB}c we highlight in red the areas of our model where a
CAP is used, indicating with a yellow line the points where $f(r)$ diverges from both
sides. Notice how CAP is set \emph{separately} on front- and back-side of the
pinhole, labeled A$^+$ and A$^-$, respectively.

\subsection{Magnetic field}

A transverse magnetic field in a graphene device can be included in the off-diagonal
elements of the Hamiltonian via Peierls substitution \cite{Luttinger1951,
	Pedersen2011}. We can formally write this as an additive term to the unperturbed
Hamiltonian,
\begin{equation}
\delta\mathbf H_{B}=
\mathbf H_0 \cdot e^{i \phi (\mathbf R, \mathbf R')} -\mathbf H_0
\end{equation}
where the phase factor is multiplied on each matrix element of $\mathbf H_0$, in a
coordinate system where the $x$ axis is aligned with graphene armchair direction. The
phase can be written as
\begin{equation}
\phi(\mathbf R, \mathbf R')=\frac{\pi}{2}\frac{B}{\Phi_0}(x+x')(y'-y),
\end{equation}
with $\Phi_0=2.07 \cdot 10^5 \, T \cdot \angstrom^2$ being the quantum magnetic flux. 

We implement this approach in \sisl/\tbtrans\ and as a benchmark we have compared to the
popular quantum transport code \kwant \cite{Groth2014}, finding good agreement.

%We implement this approach in \sisl/\tbtrans\ and as a benchmark in \Figref{sislvskwant}
%we compare to the results of the popular quantum transport code \kwant. We show that we can
%reproduce almost perfectly the effect of magnetic field on the transmission across a
%square lattice constriction, reported as one of the examples in the \kwant\ seminal paper
%\cite{Groth2014}.
%\begin{figure}
%  \centering
%  \includegraphics[width=0.8\textwidth]{figures/sisl_vs_kwant.png}
%  \caption{Comparison between \sisl-\tbtrans\ and \kwant\ with regard to the implementation
%      of magnetic field in tight-binding models. We consider the square lattice with the
%      constriction shown in the inset and calculate transmission across it at various
%      magnetic fields, confirming the reliability of our \sisl-\tbtrans\ implementation of
%      the Peierls substitution. Geometry and tight-binding parameters are taken from
%      \cite{Groth2014}. Magnetic field is represented in terms of inverse magnetic
%      flux. No field is present in the electrodes and there is no disorder in the
%      device.}
%  \label{sislvskwant}
%\end{figure}
%\NP{Although this figure is nice, I don't think it is that interesting for the
%    article... I think we can simply leave it out and state that comparison with kwant is consistent?}

\subsection{Visualization of bond-currents}

Bond currents allow imaging of spatial profiles of nonequilibrium charge and current
densities in solid-state\cite{Datta2000,Nonoyama1998} as well as molecular-scale
systems\cite{Solomon2010}. They represent local current flowing in the inter-atomic bonds
and are defined as the sum over all orbital (indices $\alpha,\beta$) currents,
$J_{\alpha\beta} = \sum_{\nu\in\alpha}\sum_{\mu\in\beta} J_{\nu\mu}$.  After being
generalized for use in honeycomb lattices \cite{Zarbo2007} direct insights could be
provided into how the massless Dirac fermions propagate between two neighboring lattice
sites in graphene and other carbon nanostructures. There is still no well established way
of visualizing bond-currents. In general they can be visualized as flow lines mapped on
the network of lattice bonds. Arrow vectors are typically used, whose thickness or length
is proportional to the magnitude of the current flowing between each pair of neighbouring
atoms \cite{Mason2013,GomesdaRocha2015,Stuyver2017,Nozaki2017}. This approach provides
useful information about both magnitude and direction of the current flowing through the
system. However if large geometries are considered it is not feasible due to the
overwhelming number of connections to be visualized. A solution to this is to visualize
bond-currents on a coarse grained grid, where only a proper average of bond-currents
within each cluster is shown as a single arrow \cite{Settnes2015}. In this case the final
result will of course depend on the cluster size and on the way the average is carried
out, but nevertheless it allows to scale the dimensions of the image at wish enabling
access to useful current maps for very large system dimensions.

\begin{figure}
	\centering
	\includegraphics[width=1.0\columnwidth]{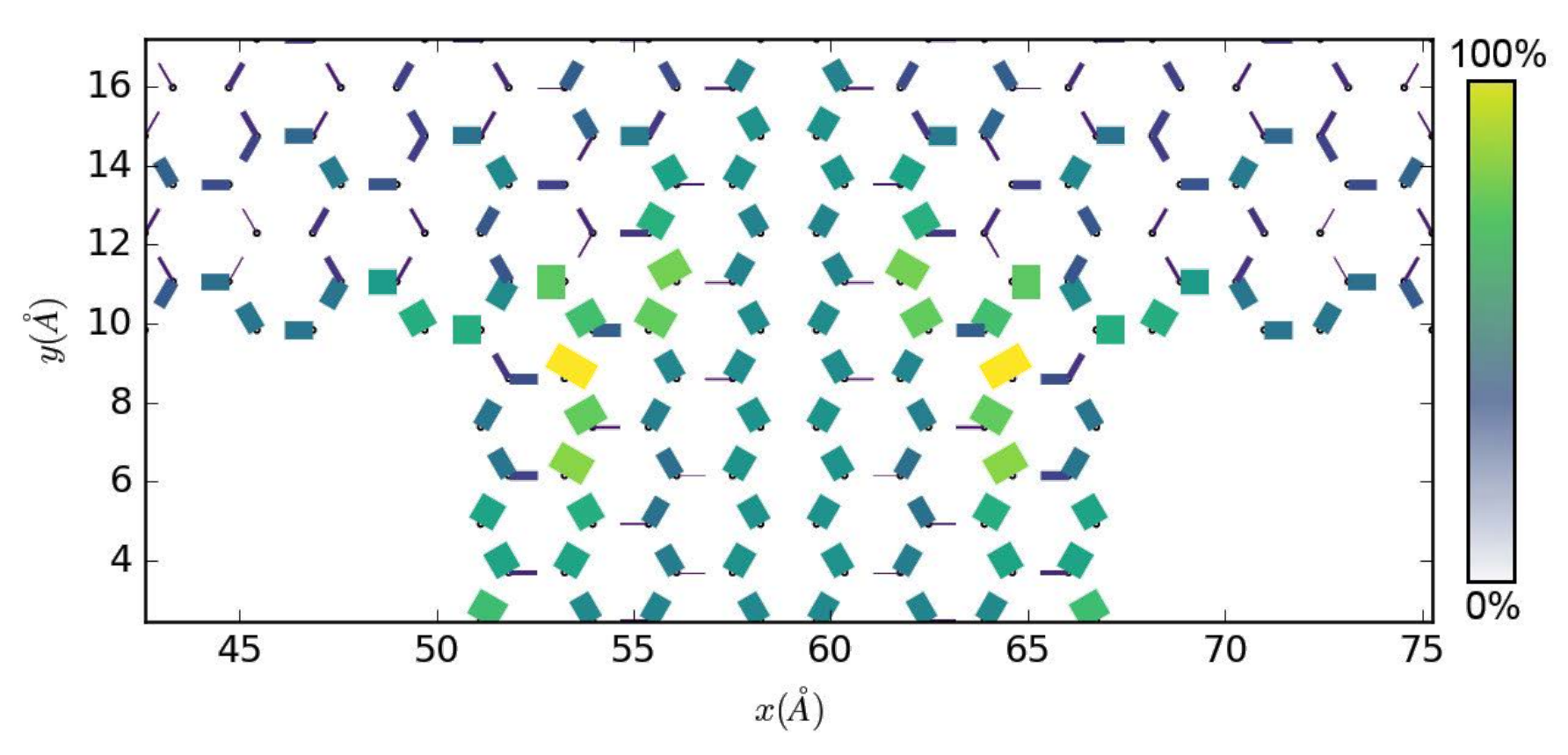}
	\caption{Positive valued bond-currents plotted as line segments of length $a_0/2$
		connecting nearest neighbor sites. Each segment is anchored to the atom from
		which the corresponding current originates. Thickness and color is scaled with
		current magnitude so that areas with low to zero current appear white.}
	\label{bcmethod}
\end{figure}

Here we visualize bond-currents as line segments of length $a_0/2$ aligned along a line
connecting each pair of neighbour sites, as illustrated in \Figref{bcmethod}. We select
only positive bond currents and for each of them we draw a segment having one extremity at
the lattice site from which the relative current originates. We scale the segment
thickness and color in proportion to the current magnitude, so that areas with low to zero
current will appear white because the bond width is reduced to zero. This approach allows
us to access graphene geometries with arbitrary large size, retaining full information
about current directionality without a need for user-defined post-processing of the
data. The same color map is adopted and color and thickness are always normalized to the
maximum value of bond-current in the system. Sometimes the color range is adjusted in
order to enhance contrast. If not stated otherwise, bond currents are always computed at
$E=E_{\mathrm F}$.

%If one is not interested in directionality, sometimes a homogeneous 2D colormap can provide a better representation of current density in the system. This can be easily obtained by summing the bond-currents originating from each site and interpolating them onto a 2D mesh grid, as shown in \Figref{bcmethod}b. For all simulations presented here the segment approach is used
%\begin{figure}
%  \centering
%  \includegraphics[width=0.9\textwidth]{figures/bcmethod.png}
%  \caption{(a) Positive valued bond-currents plotted as segments of length $a_0/2$
%      connecting nearest neighbor sites, each anchored to the atom from which the
%      corresponding current originates. Thickness and color is scaled with current
%      magnitude so that areas with low to zero current appear white. (b) A 2D interpolated
%      map of bond-currents obtained by summing all positive currents originated from each
%      atom. The interpolation mask used to create this figure has been adapted from a
%      post-processing routine implemented in \kwant.\NP{Letters on figures!}}
%  \label{bcmethod}
%\end{figure}

%\NP{Why do we then show the second one?} \GC{With Mads I remember we talked about mentioning this here, even though it is not immediately applied to the DFM, so that we could then refer to this article when the zoomIN/zoomOUT article will be out...so I made the figure. I will remove it for now..probably it's going to be more useful in the thesis ;)}

\subsubsection{Performance of \tbtrans\ and \sisl}
\begin{figure}
	\centering
	\includegraphics{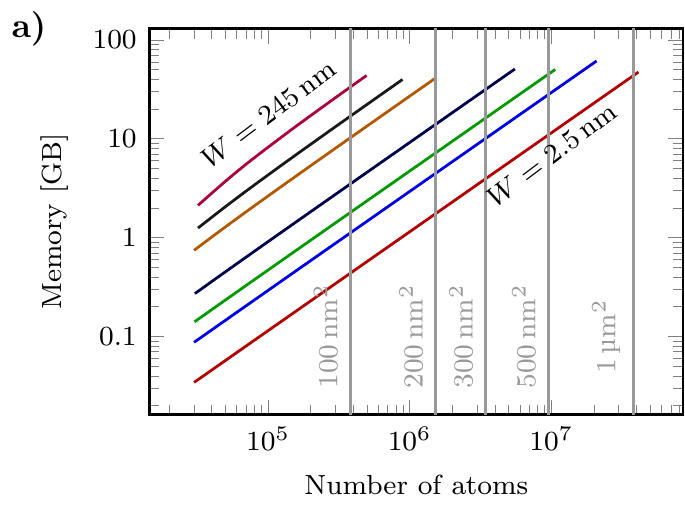}\,
	\includegraphics{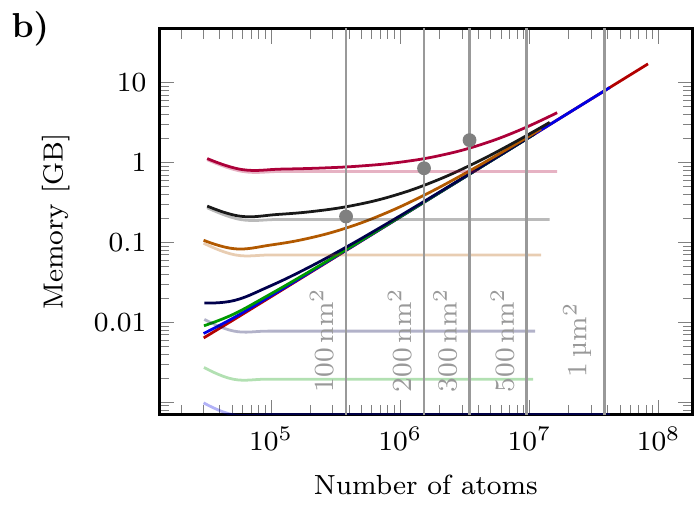}
	\caption{a) Memory requirements for varying system sizes and their memory usage when
		calculating orbital currents and orbital resolved quantities. Each line correspond
		to a different width of the system in ascending width. Increasing the width of the
		system is the main cause of memory usage since each block has to be as large as the
		orbital width. The curves terminate at around 40-50 GB because of integer
		overflows. Vertical lines indicate the number of atoms in a square unit cell of noted
		area. %
		b) Equivalent calculations as in a) while reducing the region of interest to a
		specific set of atoms. In this case the memory requirements are drastically
		reduced because the full Green's function is not needed. In this case the allowed
		range of calculated system sizes is considerably increased. The inserted dots
		indicate the memory requirements for square unit cells. The transparent lines are
		the memory used for the block tri-diagonal matrices which becomes constant. I.e. the
		memory requirements are solely determined by the sparse matrices.
	}
	\label{fig:tbtrans}
\end{figure}

The calculations are performed using \tbtrans\ which implements a highly advanced
block-tri-diagonal inversion algorithm which minimises calculations\cite{Papior2017}.
The main difficulty in calculating very large systems is the calculation of bond-currents
and orbital-resolved DOS, which requires the spectral function ($\mathbf A_i$) for a given
electrode $i$. In Fig.~\ref{fig:tbtrans} we show the memory requirements of \tbtrans\ for
the very large nearest-neighbour graphene devices with periodic boundary conditions.  In
both a) and b) vertical lines indicate system sizes of square unit cells of noted
area. Lines are ascending together with cell width, the narrowest cell being $2.5\,\nm$
and the widest $245\,\nm$ (see a).
In Fig.~\ref{fig:tbtrans}a the total memory requirements is plotted when calculating
physical quantities for the full system in one calculation. This is scaling linearly with
respect to system size. Since \tbtrans\ has been implemented with 4-byte integers, there is
an upper limit to the size of the allocated block tri-diagonal matrix. I.e. the lines
stop due to integer overflow in the code.
Fig.~\ref{fig:tbtrans}b shows memory requirements when only calculating quantities for a
selected region in the device, here chosen to include 4 lines of carbon atoms. Clearly the memory
requirements drastically reduces and becomes feasible on laptop computers. Here the gray
dots indicate the memory requirements for the square unit cells of given area. The
transparent lines represent the memory usage of the block tri-diagonal matrices, which become constant for large systems. Thus, the only memory increase is due to the
sparse matrices used to retain the Hamiltonian and bond currents etc.

\tbtrans\ is parallelized using both MPI and OpenMP enabling high-performance calculations
with large throughput. In these calculations (400.000 nearest neighbour atoms) we use 24 OpenMP threads and runs in roughly
1 minute per energy point using XeonE5-2650 machines.

A full setup, transmission calculation and post-processing using \sisl\ and \tbtrans\
takes approximately 40 minutes, with a similar amount of time spent on setting up
input and plotting.

We have shown how large scale calculations can efficiently be calculated and
post-processed. Further information about \sisl\ may be found in Ref.~\cite{Papior2017a}.

% RESULTS
\section{Results}

\begin{figure}
	\centering
	\includegraphics[width=1.0\columnwidth]{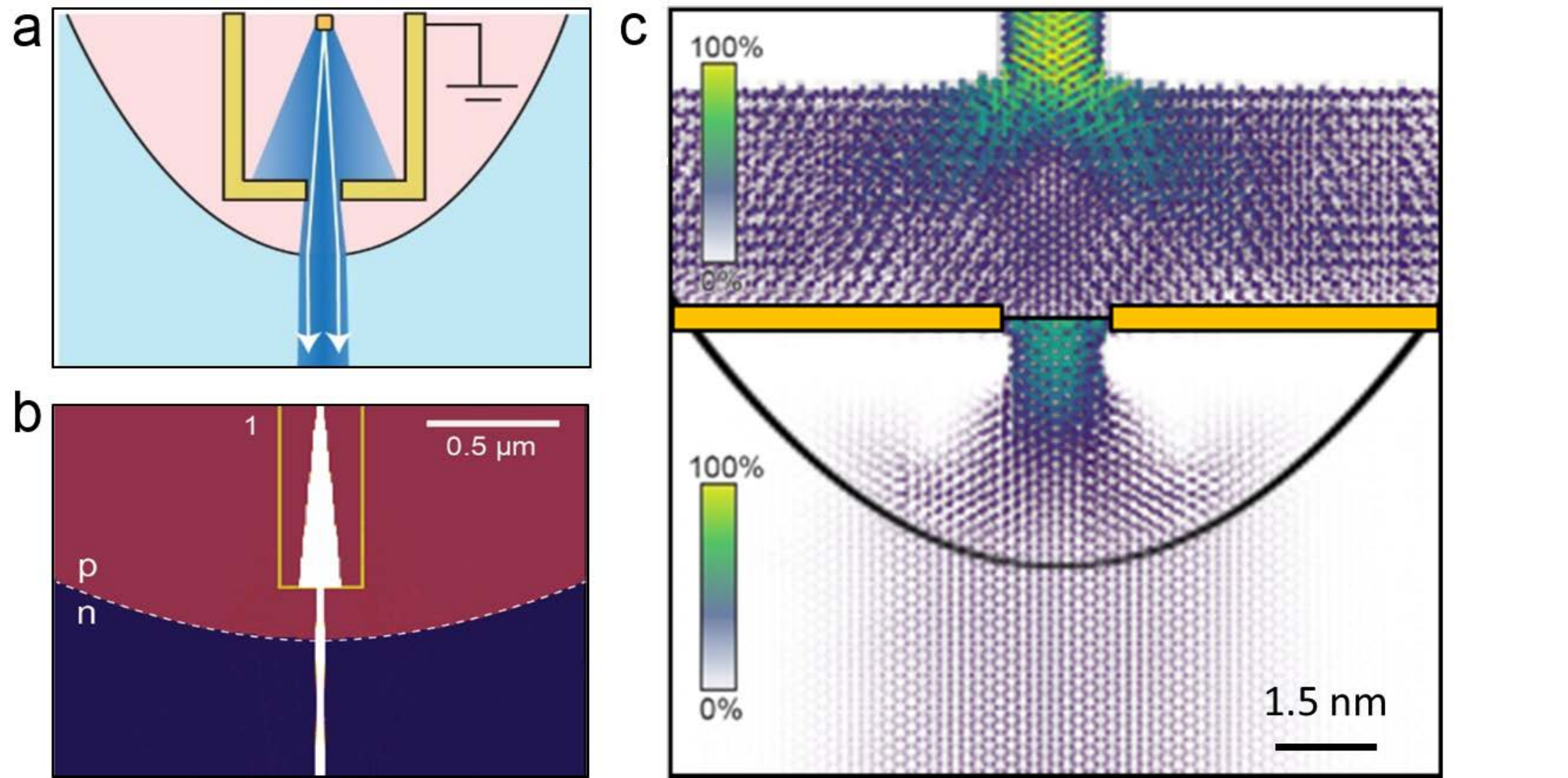}
	\caption{(a) Schematic setup of pin-hole injection with collimation $p-n$ junction.
		(b) Semiclassical calculation of schematic setup.
		(c)
		Bond-currents in proximity of the device source. Electrons are injected by a
		ribbon, filtered by an absorptive aperture and collimated by a parabolic $p-n$
		junction, in agreement with \cite{Boggild2017,Liu2017}. CAP walls are shown
		in yellow. The color scale is normalized differently before and after passing across
		the absorptive aperture to compensate for the current density loss.
		(a) and (b) are adapted from \cite{Boggild2017}.
	}
	\label{source}
\end{figure}

We use the methods presented in the previous sections to create a tight-binding model of a
DFM. In \Figref{source} we illustrate the definitive model that we use to generate
collimated beams of electrons in our calculations. Electrons are injected by a ribbon,
filtered by an absorptive pinhole and further collimated by a parabolic $p-n$ junction.  As
reported in \cite{Wang2010}, injection at the interface between the ribbon and the large
flake is largely anisotropic and can be made isotropic by using a CAP region as an absorptive
pinhole. The result is in good agreement with both the semiclassical calculations reported
in \cite{Boggild2017} and the tight-binding calculations by Liu \etal \cite{Liu2017}.

\subsection{Boundary conditions}

\begin{figure*}
	\centering
	\includegraphics[width=0.8\textwidth]{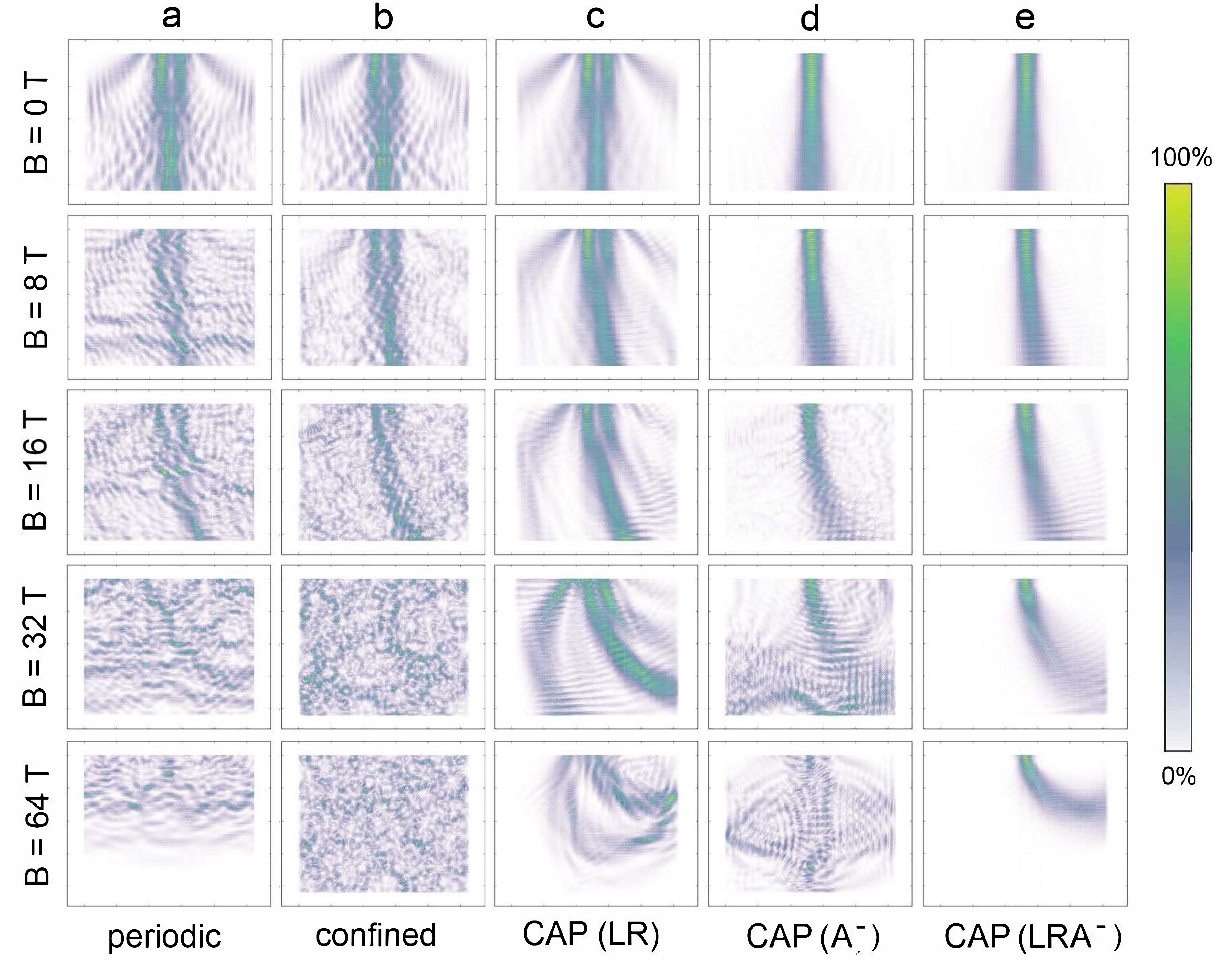}
	\caption{Bond-currents at different applied magnetic fields and various boundary
		conditions, showing a collimated beam scanning across a $100\,\nm\times100\,\nm$
		graphene cell with an energy $E_{\mathrm F}=1.0\,\ev$. Electrons are injected from the
		top side of the device, source and collimating lens are not shown so as to enhance
		beam contrast. All simulations have CAP at A$^+$ of the pinhole injector.
		(a) Periodic boundary conditions at L/R lead to a high degree of scattering, similar to (b) where hard-wall potentials on L/R
		regions confine electrons. (c) Adding CAP on L/R drastically reduces scattering
		from neighbouring cells but does not retain a collimated beam from a pin-hole
		injection. (d) Introduces the \emph{entire} pin-hole effect by absorbing
		backscattering from the $p-n$ junction.  (e) Final model with CAP on L/R and
		A$^-$. This model correctly retains a collimated beam while showing the beam in
		the limit of infinite graphene.
	}
	\label{boundary}
\end{figure*}

Choosing the correct boundary conditions is essential in reproducing the wanted
physical pictures.
The goal of the DFM example discussed below is the analysis of collimated beams injected into large graphene samples.
Here we emphasize the importance of selecting appropriate boundary conditions to
simulate such device in its limit of infinite extension.
\Figref{LRB}c is used in the following where we call L (R) the left (right) boundary of the large flake, A$^+$ the
CAP region that acts as pin-hole injector and A$^-$ the pinhole side that faces
opposite to the source. On the opposite side of injection a regular lead is placed. In
\Figref{boundary} we analyze bond-currents in the area beyond the parabolic lens at
different applied magnetic fields and various boundary conditions, progressively
switching on the CAP regions at L, R and A$^-$. In the following CAP at A$^+$ is
always used to ensure a collimated pin-hole injector.

The most straight-forward approach here would be to use periodic boundary conditions at L
and R, as considered in \Figref{boundary}a. In general this represents a very popular
choice for electronic structure calculations, where even the presence of local
perturbations in the system, e.g. defects or adsorbed molecules, can be accurately dealt
with by increasing the cell size and thus minimizing the periodic interactions. However it
is known that often periodicity is not able to correctly describe the relevant features in
non-equilibrium transport calculations \cite{Lagasse2017}. As shown in \Figref{boundary}a,
despite the very large cell used ($\approx100\,\nm$), interaction between periodic
repetitions of the source give rise to significant interference in the currents
pattern. At $B=0$ the collimated beam is still visible behind the interference fringes,
while it gets progressively suppressed as the magnetic field increases.  The situation is
somewhat similar when periodicity in L and R is replaced with hard-wall high potentials
acting as barriers (\Figref{boundary}b), except that in this case much more interference
occurs at high B fields.  An effective solution to this problem is to use CAP at L and R,
similar to what Lagasse and coauthors suggest in Ref.~\cite{Lagasse2017}. In
\Figref{boundary}c the interference is indeed reduced, especially at higher B fields,
where the beam is now clearly visible beyond the fringes.  Nevertheless one can notice
that the electron beam is still not very well collimated: already at $B=0$ it splits into
several narrow beams after crossing the parabolic lens. This is due to internal
reflections occurring between the parabolic junction and the region A$^-$ of the pinhole,
which in \Figref{boundary}c is not equipped with CAP. This is an artificial effect,
since realistic grounded electrodes contacting graphene to create apertures would at least partly \cite{Barnard2017} absorb
electrons impinging onto them from all possible directions. In \Figref{boundary}d we
demonstrate that this backscattering effect can be eliminated by switching on CAP at the
A$^-$ region. The combined application of CAP in L, R, A$^-$ eventually allows us to get
rid of most interference effects for all magnetic fields considered, see \Figref{boundary}e.

% \begin{figure}
%   \centering
%     \includegraphics[width=1.0\textwidth]{figures/boundaries_src.png}
%   \caption{{\bf ...} ({\bf a-b}) ...}
%   \label{boundary_src}
% \end{figure}

We conclude that investigating an injected beam in the limit of infinite graphene devices
requires a specific and an elaborate set of boundary conditions to filter out artificial
backscattering processes. Importantly a A$^-$-side CAP is necessary to absorb
backscattered electrons from the $p-n$ junction as well as beam-bending from the magnetic
field.

\subsection{Comparison with semiclassical simulations of DFM}

In order to gain a deeper insight on the role of quantum coherence effects in the DFM, we
consider some of the systems studied in Ref.~\cite{Boggild2017} where collimated electron
beams are focused onto circular Veselago dots (VD) of various size.  We concentrate on the
mesoscopic limit, $l_{\mathrm{mfp}}>L\gg\lambda_{\mathrm F}$, where the mean free path $l_{\mathrm{mfp}}$ of electrons is larger than
the characteristic length $L$ of the system, which in turn is much larger compared to the
Fermi wavelength $\lambda_{\mathrm F}$. In the following we consider explicitly the situation that the phase coherence length is infinite, i.e. that the system is fully coherent. 

We note that the area available for the quantum simulations is smaller than the structures
considered in realistic setups such as those in Ref.~\cite{Boggild2017}.  We deal with
this by effectively scaling the graphene bond-length while retaining the number of
atoms in our TB model via a scaling parameter $s=d_0/d$, defined as the ratio between the real diameter $d_0$ of
the \emph{full, non-scaled} VD that we would like to simulate and the \emph{actual}
diameter $d$ of the VD designed in our geometry. We assume $n_0=10^{12}\,\cm^{-2}$ to be
the electron density in the non-scaled pristine graphene system, which (in the linear band
approximation) corresponds to
$E_{\mathrm F0}=\hbar v_{\mathrm F}\sqrt{\pi n_0}=0.113\,\ev$ and
$\lambda_{\mathrm F0}=2\pi/\sqrt{\pi n}=35.4\,\nm$.  A qualitatively correct electron flow
around the VD of diameter $d$ in our bond-currents calculations can thus be captured by
simply dividing the Fermi wavelength by the scaling factor $s$, yielding
$\lambda_{\mathrm F}=\lambda_{\mathrm F0}/s$ and $E_{\mathrm F}=s\, E_{\mathrm F0}$. The
key step of the scaling procedure is to keep the diameter vs. Fermi-wavelength ratio
constant \cite{Heinisch2013, Caridad2016, Jiang2017}. This approach can be thought as a
particular case of the more general scaling method presented in \cite{Beconcini2016},
hence we refer readers to this for further details.

\begin{figure}
	\centering
	\includegraphics[width=1.0\columnwidth]{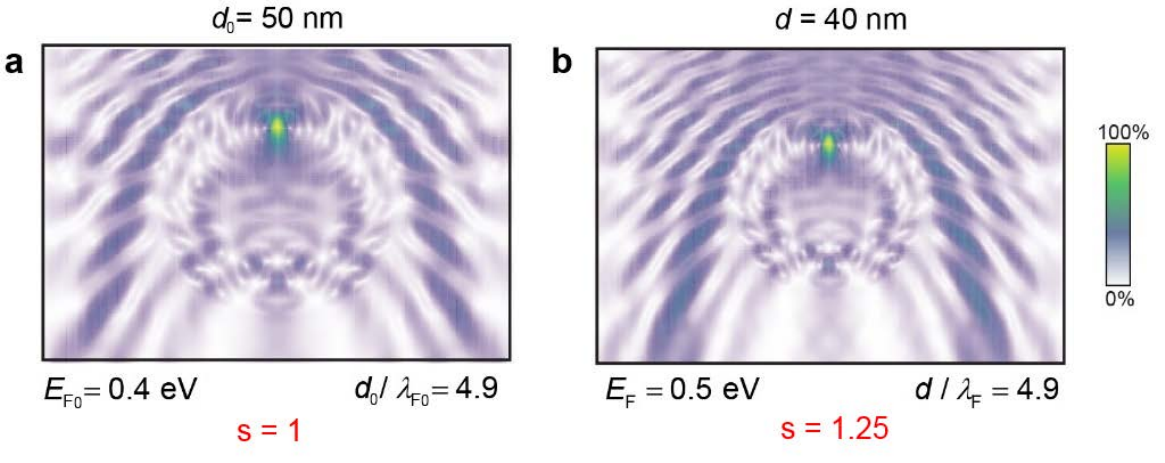}
	\caption{Comparison of bond-current pattern of two VDs with different size and energy
		but same diameter vs. Fermi-wavelength $d/\lambda_{\mathrm F}=4.9$. Both maps show a
		$100\,\nm \times 75\,\nm$ area of the device in front of the collimation area. The
		scaling method is used in (b) showing that the electron flow inside and outside the
		VD in (a) can be qualitatively reproduced by using appropriate scaling
		parameters.}
	\label{NatComm1}
\end{figure}

Our scaling method is exemplified in \Figref{NatComm1}. Let us assume that the VD on the
left with $d_0=50\,\nm$ is the original \emph{non-scaled} VD that we want to study
($s=1$). In \Figref{NatComm1}b we demonstrate that by applying the scaling
procedure with $s=1.25$ we are able to reduce the VD size and produce bond-currents inside
and outside the VD which are indistinguishable from \Figref{NatComm1}a. In general,
this approach enables us to effectively simulate systems that are $s$ times larger
than the actual $100\,\nm \times 100\,\nm$ geometry considered in our
calculations. Using $s=2$, for example, is equivalent to analyze a
$200\,\nm \times 200\,\nm$ graphene system with an equivalent number of atoms.

%The applicability of this scaling method is shown in \Figref{NatComm1} where we illustrate
%the caustic bond-current patterns for two VD with diameters $d=50\,\nm$ and $d=40\,\nm$,
%and $E_{\mathrm F}=0.4\,\ev$ and $0.5\,\ev$, respectively, both with
%$d/\lambda_{\mathrm F}=4.9$, corresponding to the scaled diameter
%$d_{\mathrm{scaled}}=178\,\nm$. It is easy to see that the bond-current patterns are
%virtually indistinguishable, thus supporting our approach of rescaling the calculations.
%\begin{figure}
%	\centering
%	\includegraphics[width=0.8\textwidth]{figures/NatComm_1.png}
%	\caption{Comparison of bond-current caustic pattern of two VDs with different size and
%		energy. The scaling method is used in both cases and show that the same bond-current
%		map can be obtained by determining correct scaling parameters. In this case both
%		VD's have $d/\lambda_{\mathrm F}=4.9$.}
%	\label{NatComm1}
%\end{figure}

In the following we will always indicate above every figure the \emph{full, non-scaled}
diameter $d_0$ of the VD, while below we will provide the diameter vs. Fermi-wavelength
ratio $d/\lambda_{\mathrm F}$ considered and the value of $s$ used to scale the system
down to our $\sim 400.000$-orbitals TB model.

\begin{figure}
	\centering
	\includegraphics[width=1.0\columnwidth]{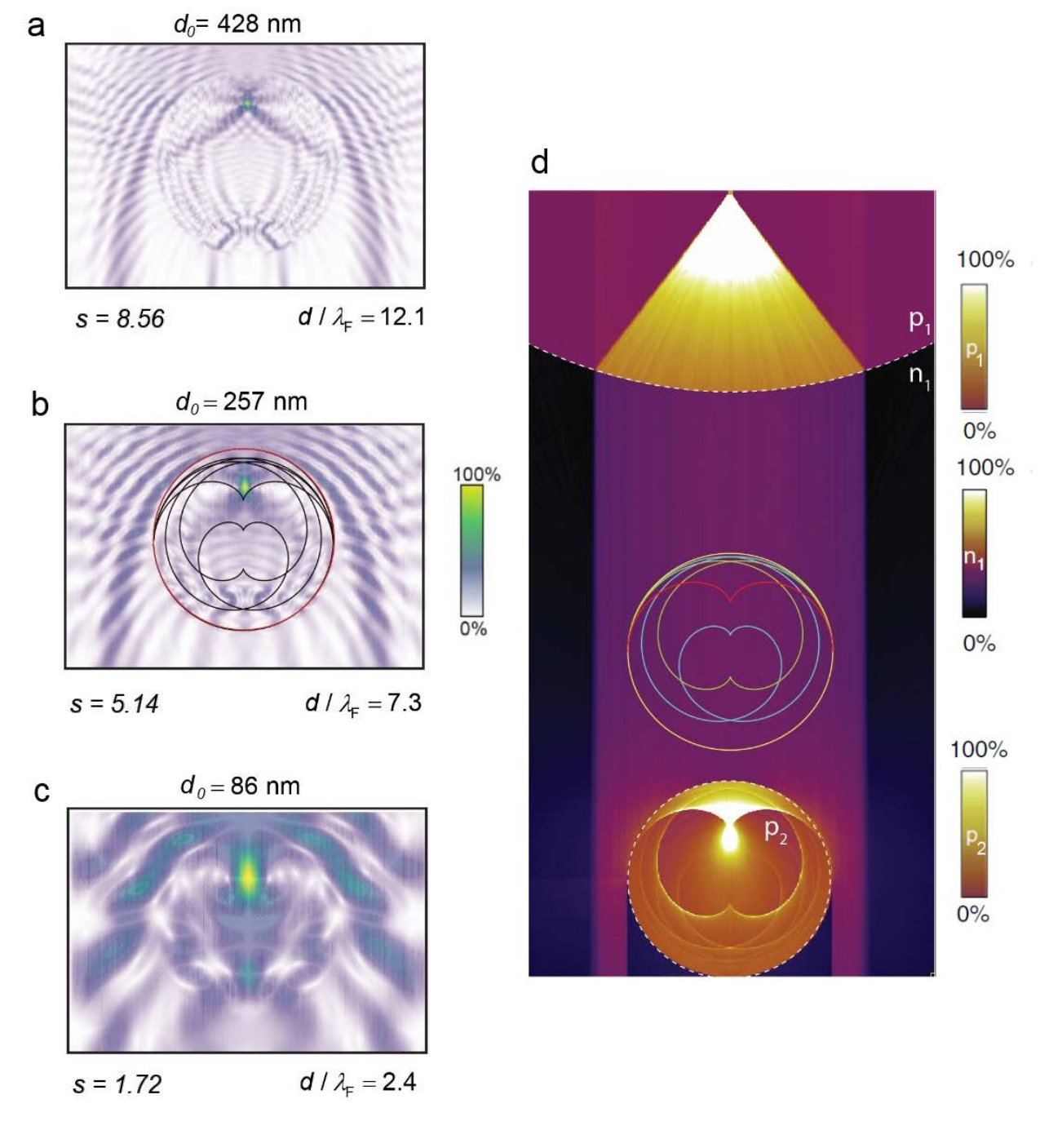}
	\caption{Bond current caustic patterns around VD with (a) $d_0=428\,\nm$, (b)
		$257\,\nm$ and (c) $86\,\nm$ non-scaled diameter, in comparison with (d)
		current density obtained in Ref.~\cite{Boggild2017} via semiclassical
		simulations. The classical caustics from Ref.~\cite{Cserti2007} are superposed
		in (b) and (d). The source and collimation area comprising ribbon emitter,
		absorptive pinhole and parabolic lens is not shown.}
	\label{NatComm2a}
\end{figure}

Figure~\ref{NatComm2a} shows caustic patterns inside VD with different scaled diameters,
namely $d_0=428\,\nm$, $257\,\nm$ and $86\,\nm$, generated by scattering of a wide beam of
electrons collimated with a $f\approx 10\,\nm$ parabolic lens in front of a $1.5\,\nm$
aperture. These diameters correspond to scaling parameters $s=8.56$, $5.14$ and $1.72$,
respectively, allowing us to compare to some of the systems studied by semi-classical
simulations in Ref.~\cite{Boggild2017}. We focus on the area inside the dot and observe
characteristic caustic patterns at all energies which are in reasonable agreement with the
results reported in Ref.~\cite{Boggild2017} and by Agrawal
\etal\cite{AgrawalGarg2014}. This is especially true with regard to the position of the
main cusp in the first caustic line.
A better description of the higher order caustics while keeping Fermi energies ($<1\,\ev$)
within the limits of the linear band approximation would require creation of VD with
scaled diameters at least 2--3 times larger than the ones considered here, i.e.
$d\ge 200 \,\nm$, which is not possible within our $100\,\nm \times100\,\nm$ graphene
flake.
%\NP{It is unclear to the reader how large they actually are? Is the full system *always* 400.000 orbitals?} %\GC{Yes, the system has always 400.000 orbitals. Maybe the way I write it now is more clear..}. 
For similar dot dimensions quantum mechanical calculations using plane waves have indeed proven to reproduce reliable optical geometrical features such as peaks in forward scattering \cite{Caridad2016}. 
%\begin{figure}
%  \centering
%  \includegraphics[width=0.8\textwidth]{figures/NatComm_2.pdf}
%  \caption{Bond current caustic patterns for a $50\,\nm$ VD with (a) $E_{\mathrm F}=1.0\,\ev$,
%      (b) $0.6\,\ev$ and (c) $0.2\,\ev$. These energies correspond to VDs with diameters
%      $d_{\mathrm{scaled}} = 428\, \nm$, $257\, \nm$ and $86\, \nm$ for $n=10^{12} \,\cm^2$
%      ($E_{\mathrm F}=0.117\,\ev$), which compare well with (d) the current density considered
%      in Ref.~\cite{Boggild2017}. The classical caustics from Ref.~\cite{Cserti2007} are
%      superposed in (b) and (d). The source area comprising ribbon emitter, absorptive
%      pinhole and parabolic lens is not shown in the figure.}
%  \label{NatComm2}
%\end{figure}

\begin{figure}
	\centering
	\includegraphics[width=1.0\columnwidth]{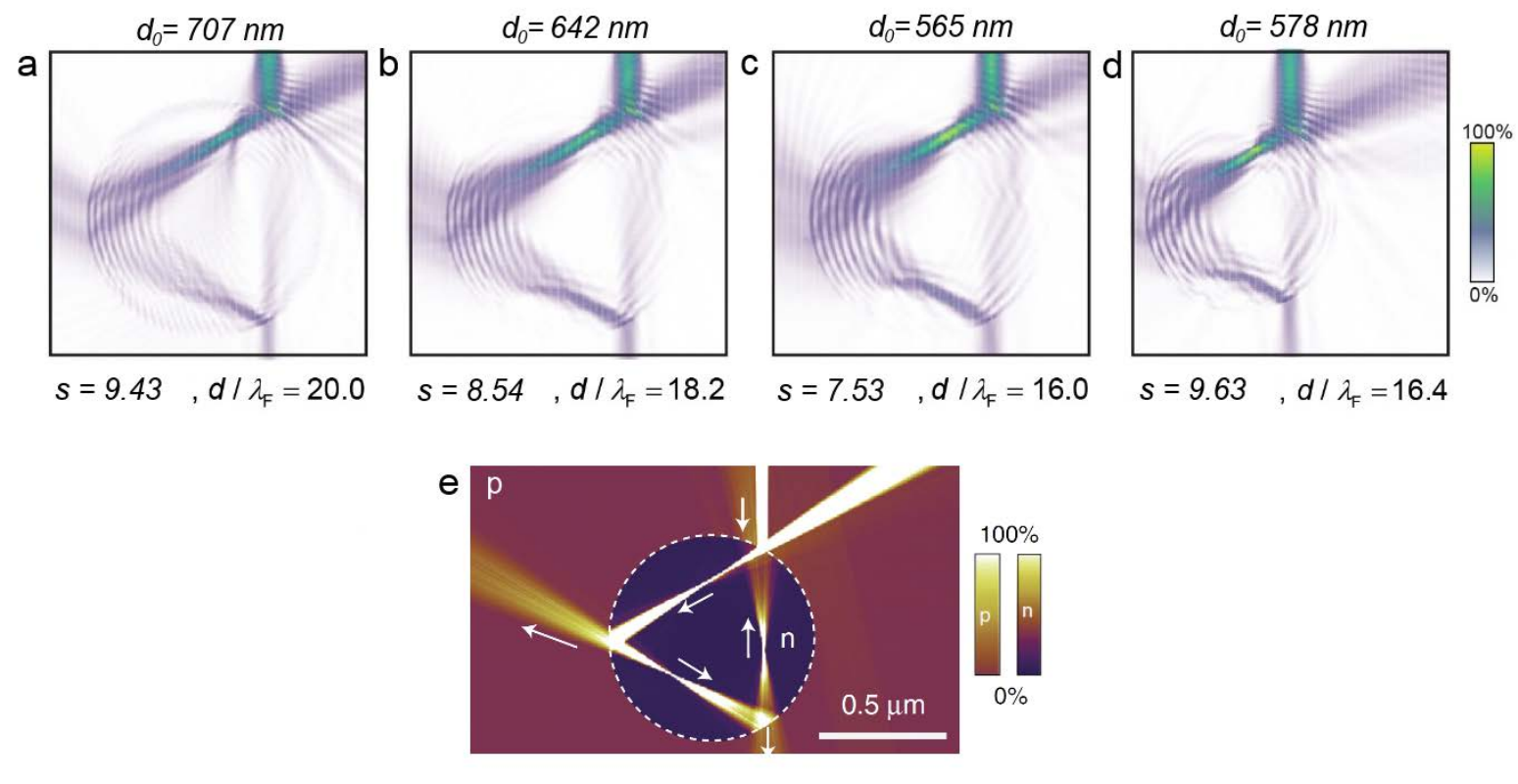}
	\caption{A focused DF beam is impinging on a large circular VD with different non-scaled
		diameters $d_0$ and similar ratio $d/\lambda_{\mathrm F}$. The bond currents for
		$d_0 = 707\,\nm$ is in agreement with (e) the semiclassical current density of the
		$700\,\nm$ dot diameter from \cite{Boggild2017}. Comparison of (a-d) confirms that
		structures with similar $d/\lambda_{\mathrm F}\approx16$ have very similar bond
		current distribution.}
	\label{NatComm3}
\end{figure}

Figure \ref{NatComm3} shows a sharper electron beam, focused and collimated by combining a
$1.5\,\nm$ aperture and a $f=4\,\nm$ parabolic lens, scattering on a VD with various
non-scaled diameters $d_0$ and similar ratio $d/\lambda_{\mathrm F}$.  In particular
Figure~\ref{NatComm3}a has the same scaled diameter $d_0=707\,\nm$ as the VD simulated in
many of the structures considered in \cite{Boggild2017}, thus comparing well both
with respect to the emission jets and the internal, polygonal current
resonances. Clearly interference effects are of higher importance in the quantum
mechanical calculations, which are seen as beam broadening inside and outside the VD.
%\begin{figure}
%	\centering
%	\includegraphics[width=0.8\textwidth]{figures/NatComm_3.png}
%	\caption{A focused DF beam is impinging on a large circular VD with a diameter of
%		$75\,\nm$ and energies $E_{\mathrm F}=1.1\,\ev$, $1.0\,\ev$ and $0.9\,\ev$. The bond
%		currents for $E_{\mathrm F}=1.1\,\ev$ ($d_{\rm scaled} = 707\,\nm$) is in agreement with
%		(e) the semiclassical current density of the $700\,\nm$ dot diameter from
%		\cite{Boggild2017}. Comparison of (c) and (d) confirms that structures with similar
%		$d/\lambda_{\mathrm F}\approx16$ have very similar bond current distribution.}
%	\label{NatComm3}
%\end{figure}

\begin{figure}
	\centering
	\includegraphics[width=1.0\columnwidth]{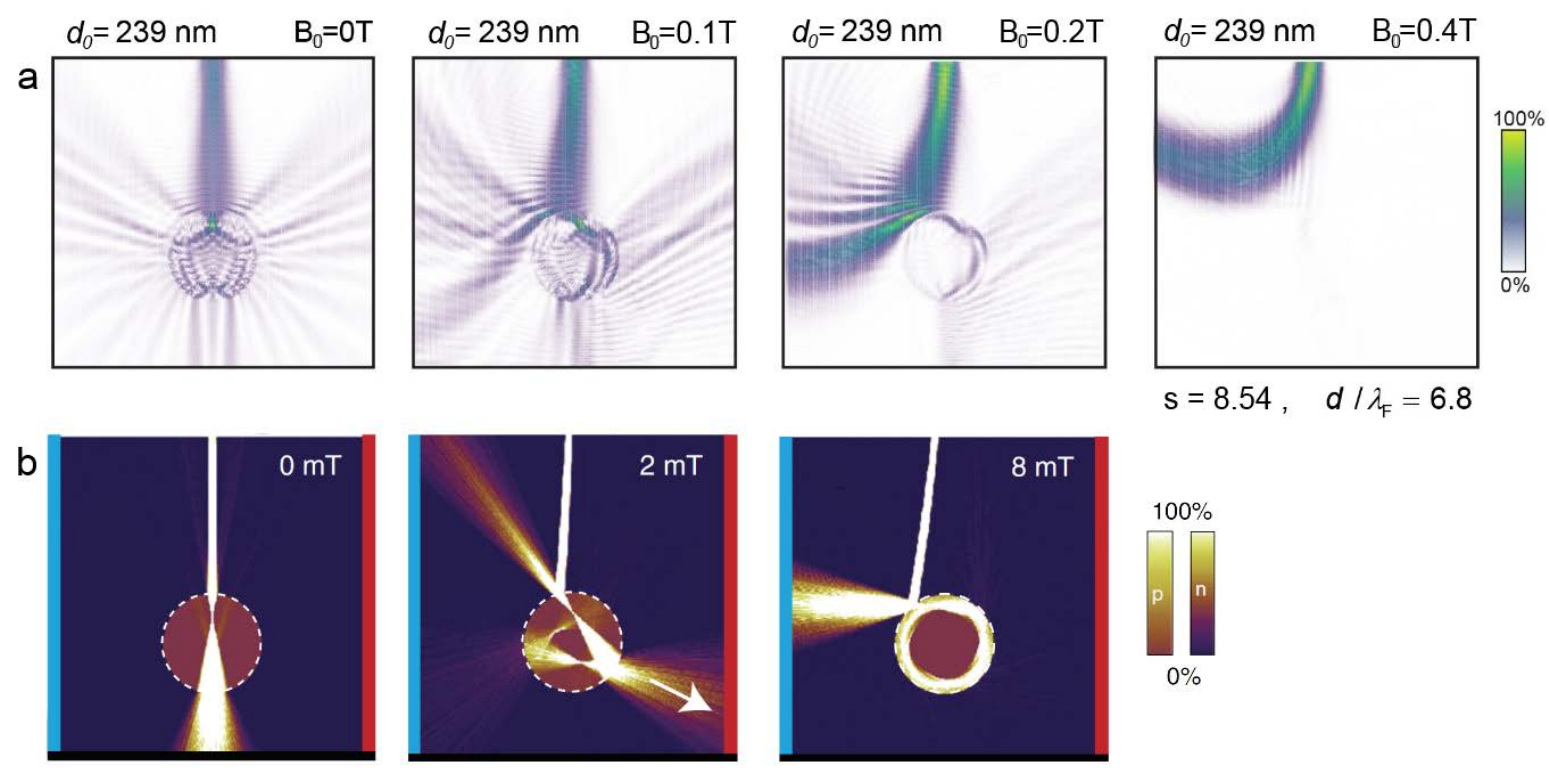}
	\caption{Collimated DF beam scanning across a small VD with non-scaled diameter
		$d_0=239\,\nm$ and ratio $d/\lambda_{\rm F}=6.8$. The bond current scattering
		patterns in (a) roughly resemble the semiclassical simulations from
		Ref. \cite{Boggild2017}, shown in (b). The scaled values of magnetic field used in our model 
		are $B=0\,\mathrm{T}$, $8\,\mathrm{T}$, $16\,\mathrm{T}$ and $32\,\mathrm{T}$.}
	\label{NatComm4}
\end{figure}

In \Figref{NatComm4} we show the bond-currents obtained by scanning the beam using
different magnetic fields, ranging from $B=0\,\mathrm T$ to $B=32\,\mathrm T$. Also in this case
we find a good qualitative agreement with the semi-classical studies\cite{Boggild2017},
capturing both emitted jets and whispering channels within the circular $p-n$ junctions.
The beam size difference as well as the interference in- and outside the VD are the main
differences that amount to the leaking of electrons in the VD. 
We point out that when we model a large physical system such as that in \Figref{NatComm4}b with a scaled-down model we have to scale the magnetic field such that the flux is the same. This is done by using the relation 
$B=s^2\,B_0$, where $B_0$ is the original non-scaled magnetic field \cite{Beconcini2016}. 
In addition to this scaling we make use of larger magnetic fields compared to the semiclassical simulations in Ref. \cite{Boggild2017}, 
in order to access beam scattering off-axis with respect to the VD center. By doing this we compensate the small source/VD distance in our $100\,\nm \times 100\,\nm$ geometry, which could not be further extended due to limits of applicability of the scaling procedure (e.g. see Fig.2 in \cite{Beconcini2016}).

%\begin{figure}
%	\centering
%	\includegraphics[width=0.8\textwidth]{figures/NatComm_4.png}
%	\caption{Collimated DF beam scanning across a small VD ($d=28\,\nm$,
%		$d_{\rm scaled} =239\,\nm$) with an energy $E_{\rm F}=1.0\,\ev$. The bond current
%		scattering patterns roughly resemble the semiclassical simulations reported in
%		\cite{Boggild2017}}
%	\label{NatComm4}
%\end{figure}

%\begin{figure}
%	\centering
%       \includegraphics[height=0.5\textheight]{figures/{Bscanw2nmf10nmaperture1.5nmdot28nm}.png}
%	\caption{Magnetic field dependent behaviour of caustic patterns in a VD with $d=28\,\nm$, $d_{\rm scaled} =239\,\nm$), obtained using broad electron beams with energy $E_{\rm F}=1.0\,\ev$ obtained using a parabolic lens with focal length $f=10\, \nm$.}
%	\label{NatComm2b}
%\end{figure}

\section{Conclusions}

In conclusion, we carry out large-scale quantum transport calculations based on simple
tight-binding models of graphene and the non-equilibrium Green's function method. We
report on how to include the effects of $p-n$ junctions, magnetic field and complex
absorptive potentials into the calculations, as simple perturbative terms to the Hamiltonian. We show how different choices of boundary conditions lead to different current
features in the system, setting up local CAP regions in order to minimize artificial
interference in the current patterns. We reproduce, from a fully atomistic perspective, some
key features of electron transport in a DFM, such as electron beam collimation, deflection
and scattering off circular Veselago dots, presenting a direct comparison with the
semi-classical results reported in \cite{Boggild2017}. As expected, the quantum transport
simulations show that current density of structures, which are large compared to the Fermi
wavelength, show reasonable resemblance with the classical calculations. On the other hand,
it is evident that quantum coherence leads to bond current patterns with richer emission
and reflection structures, which may be utilized to extract more detailed information of
targets than possible with semi-classical calculations.

\section{Acknowledgements}
We are thankful to Dr. Jose Caridad for discussions.  We acknowledge funding from Villum
Fonden (grant no. 00013340) and the Danish research council (grant no. 4184-00030). The
Center for Nanostructured Graphene (CNG) is sponsored by the Danish Research Foundation,
Project DNRF103.

\section*{References}
\bibliographystyle{unsrt}
\bibliography{largeTB_v9_ARXIV}

\end{document}